\documentclass[sigchi]{acmart}
\AtBeginDocument{%
  \providecommand\BibTeX{{%
    \normalfont B\kern-0.5em{\scshape i\kern-0.25em b}\kern-0.8em\TeX}}}

\setcopyright{acmcopyright}
\copyrightyear{2021}
\acmYear{2021}
\acmDOI{10.1145/1122445.1122456}

\acmConference[C\&C '23]{C\&C '23: ACM SIGCHI Conference on Creativity and Cognition}{June 13--17, 2022}{GatherTown}
\acmBooktitle{C\&C '22: ACM SIGCHI Conference on Designing Interactive Systems,
  June 13--17, 2022, GatherTown}
\acmPrice{15.00}
\acmISBN{XXX}

\copyrightyear{2023}
\acmYear{2023}
\setcopyright{acmlicensed}\acmConference[C\&C '23]{Creativity and Cognition}{June 19--21, 2023}{GatherTown}
\acmBooktitle{Creativity and Cognition (C\&C '23), June 19--21, 2023, GatherTown}
\acmPrice{15.00}
\acmDOI{10.1145/3591196.3593337}
\acmISBN{979-8-4007-0180-1/23/06}

\newcommand{\fb}[1]{#1}

\title[Freeform Templates]
{Freeform Templates: Combining Freeform Curation\\with Structured Templates
}

\author{Stephen MacNeil}
\email{stephen.macneil@temple.edu}
\affiliation{%
  \institution{Temple University} 
  \streetaddress{1801 N Broad St}
  \city{Philadelphia}
  \state{PA}
  \country{USA}
  \postcode{19122}
}

\author{Ziheng Huang}
\email{z8huang@ucsd.edu}
\affiliation{%
  \institution{University of California, San Diego}
  \streetaddress{9500 Gilman Drive}
  \city{La Jolla}
  \state{CA}
  \country{USA}
  \postcode{92093}
}

\author{Zijian Ding}
\email{ding@umd.edu}
\affiliation{%
  \institution{University of Maryland College Park}
  \city{West Hyattsville}
  \state{MD}
  \country{USA}
}

\author{Kenneth Chen}
\email{kechen@gatech.edu} 
\affiliation{%
  \institution{Georgia Tech University}
  \city{Atlanta}
  \state{GA}
  \country{USA}
}

\author{Alexander Yu}
\email{awy001@ucsd.edu}
\affiliation{%
  \institution{University of California, San Diego}
  \streetaddress{9500 Gilman Drive}
  \city{La Jolla}
  \state{CA}
  \country{USA}
  \postcode{92093}
}

\author{Kendall Nakai}
\email{kendallnakai@gmail.com}
\affiliation{%
  \institution{Microsoft}
  \city{Redmond}
  \state{WA}
  \country{USA}
}

\author{Steven P. Dow}
\email{spdow@ucsd.edu}
\affiliation{%
  \institution{University of California, San Diego}
  \streetaddress{9500 Gilman Drive}
  \city{La Jolla}
  \state{CA}
  \country{USA}
  \postcode{92093}
}

\renewcommand{\shortauthors}{MacNeil, et al.}

\begin{abstract}

Online whiteboards are becoming a popular way to facilitate collaborative design work, providing a free-form environment to curate ideas. However, as templates are increasingly being used to scaffold contributions from non-experts designers, it is crucial to understand their impact on the creative process. In this paper, we present the results from a study with 114 students in a large introductory design course. Our results confirm prior findings that templates benefit students by providing a starting point, a shared process, and the ability to access their own work from previous steps. While prior research has criticized templates for being too rigid, we discovered that using templates within a free-form environment resulted in visual patterns of free-form curation where concepts were spatially organized, clustered, color-coded, and connected using arrows and lines. We introduce the concept of \textit{`Free-form Templates'} to illustrate how templates and free-form curation can be synergistic.

\end{abstract}


\begin{document}

\maketitle

\section{Introduction}

Collaborative digital whiteboard platforms---like Miro, FigJam, Mural, and Google JamBoard---are becoming a popular way to facilitate collaborative design work in online settings. These digital whiteboards are being used by both professional design teams and also by non-designers in classroom settings~\cite{linder2015beyond, Lupfer2019multiscaledesign} and community design workshops~\cite{yarmand2021facilitating} to coordinate design activity. Prior work by creativity researchers has focused on how whiteboards can support creative work by providing a flexible and unlimited canvas for users to organize and continually reinterpret their ideas~\cite{kerne2017strategies, lupfer2016ideamache, lupfer2016patterns}. These behaviors, known as free-form curation, can result in spontaneous, improvisational, and divergent creative processes~\cite{kerne2017strategies, lupfer2016ideamache, lupfer2016patterns}.

Although digital whiteboards facilitate creative design work and make it easier for individuals to collaborate online, they largely replicate features of existing physical whiteboards. One of the unique affordances offered by digital whiteboards is that templates can be used to scaffold design work. Templates have been shown to help novices in a variety of design tasks such as writing problem statements~\cite{macneil2021framing}, crafting emails~\cite{hui2018introassist}, and even designing experiments~\cite{pandey2018docent}. For these reasons, workshop facilitators often use or adapt `activity templates' to scaffold the design contributions of workshop participants. While these templates can help novices perform more like experts, they are often criticized for being overly restrictive~\cite{macneil2021framing, hui2018introassist, bauer2013designlibs}. This raises the question of whether there may be ways to provide the scaffolding benefits of templates while mitigating their potential to stifle creative expression.  


In this paper, we investigate how novices engaged with templates in a free-form digital whiteboard environment through a mixed-methods study of how students used online whiteboards in a large design course. In our study, design students (N=114) used an activity template to complete their coursework. The activity template provided students with instructions and input areas to guide students' contributions. The input areas took the form of dotted boxes which varied in size from small to large with the intention to guide students to know how much to contribute. We evaluated students' processes and outcomes through a thematic analysis of interview data and a content analysis of each team's whiteboard. 

Our results suggest that students used the templates extensively as a starting point to overcome the `blank page', to establish 'common ground' for coordinating design activity, and to provide an implicit cue for students to reflect on their prior work. Based on our content analysis, we also observed `patterns of free-form curation behaviors'~\cite{kerne2017strategies, lupfer2016patterns} both inside and outside the template. For instance, students used the large space within the input areas to further organize, annotate, and relate their ideas. Similarly, they used the space outside the templates to create additional flexibility and to think beyond the planned activities. 

Based on these results, we introduce the concept of \textit{free-form templates} which describes how the flexibility provided by digital whiteboards can help to address the overly restrictive nature of traditional templates~\cite{macneil2021framing, hui2018introassist, bauer2013designlibs} while even affording opportunities for users to add additional structure. We discuss the implications for the design of future template-based creativity support tools which may shed the confines of traditional interfaces and embrace the flexibility provided by open canvases.

This paper makes the following contributions:
\begin{itemize} 
    \item A study of how design students use templates within a free-form digital whiteboard environment with insights about how templates provide students with a starting point, a process to follow, and serve as a common ground to coordinate design work. 
    \item Evidence of free-form curation behaviors both inside and outside the provided templates which suggests that {`Free-form Templates'} can provide students with both structure guidance and flexibility to curate their work.
    
    
\end{itemize}





\section{Background} 

Design workshops are a common method for engaging groups of people in collaborative design activity. These workshops are often in-person events, but digital whiteboards are making these rich collaborative design experiences possible even online. Our work builds large-scale participatory efforts by using digital whiteboards and templates to scaffold a creative design process.

\subsection{Digital whiteboards support collaborative design in online environments} 

Designers and design teams traditionally use physical whiteboards to brainstorm, develop, sketch, and organize ideas~\cite{inie2020interaction, tang2009whiteboard, geyer2012ideavis, bellamy2011sketching, mangano2014supporting}. Designers use whiteboards to externalize their ideas through writing with markers, attaching sticky notes, or sketching. Whiteboards provide designers with much needed flexibility as content can be easily added, removed, rearranged, and grouped with erasers and markers~\cite{everitt2003twoworlds}. Whiteboards are especially helpful for coordinating design activities between team members by serving as a `boundary object'~\cite{bodker1994scenarios, star1989institutional} that aids communication and recollection of work-in-progress. By navigating these whiteboards, designers can reinterpret others' contributions as `found objects' to take an idea, give it a new context, and therefore new significance~\cite{lippard1971dadas, lupfer2016ideamache}. 

In online settings, digital whiteboards largely replicate the affordances offered by physical whiteboards. Designers can add sticky notes, drawings, and images to a large two-dimensional canvas which build on decades of previous research on zoomable interfaces such as PAD++~\cite{bederson1994pad}, IdeaMache~\cite{kerne2017strategies, lupfer2016patterns}, and LiquidText~\cite{tashman2011liquidtext}. However, unlike physical whiteboards which can be erased and altered, digital whiteboards support tracking changes, versioning, and sharing~\cite{gumienny2013supporting, singh2020workshops, 2010branhamreboard}. This ability to save and share digital whiteboards allows designers to easily document their design process and return to previous versions of their work. Digital whiteboard platforms like Miro~\footnote{https://www.miro.com}, FigJam~\footnote{https://www.figma.com/figjam/}, and Mural~\footnote{https://mural.com} additionally support collaboration so that a small or extremely large group of designers can work on different representations of a design at the same time. When working synchronously, each designer's cursor is visible to all other collaborators---like a digital body language. This peripheral awareness of what others are doing makes it possible to tell what others are doing and where they are focusing their attention. Though people might have better awareness of peripheral design activity in-person, digital whiteboards can scale activity to a much larger number of synchronous or asynchronous collaborators.


%

\subsection{Digital whiteboards support free-form curation} 





Curation is a process of deliberately selecting and organizing information or objects to communicate a specific insight or perspective~\cite{wolff2013curation}. Curation is often considered in the context of art and museum exhibits, where expert curators tell stories through the careful selection of artifacts. 
Designers make sense of complex design problems by curating their design artifacts and ideas using tools like concept maps~\cite{novak2006theory}, affinity diagrams~\cite{harboe2015real}, personas, and priority matrices~\cite{grudin2002personas}. Through this process of organizing and reorganizing their design ideas, designers are able to identify creative connections, make `creative leaps', and refine their understanding of the design situation.

Digital whiteboards offer a flexible two-dimensional canvas on which designers can engage in sense-making through curating design ideas and design artifacts. Linder, Kerne, and their research team coined the term `free-form curation' to describe the process where concepts are gathered, assembled, annotated, and exhibited on a digital whiteboard to form a visual semantic whole ~\cite{linder2015beyond, kerne2017strategies}. This work was later expanded to show how the free-form curation process helped students to discover relationships between concepts~\cite{kerne2017strategies} and identified visual patterns of curation that emerged organically when curators organized their concepts through a free-form curation process~\cite{lupfer2016patterns}. For example, the spatial arrangement (morphology) and grouping of elements (group) may cause curators to creatively explore and associate concepts to form new understanding. The unrestricted path through which curators reason about elements may lead to divergent wanderings to unexpected spaces (path). The layering of information may help curators to create visual hierarchy and curate at different levels of abstraction (overlap).


These visual organizations of elements can lead to new ideas and associations, but they rely on chance and on the curator’s ability to form associations, create levels of abstraction, and uncover the path of exploration. These free-form spaces not only rely on the curator's ability to iteratively make sense of what they see but also require curators to put concepts on the canvas to begin with. Filling out a blank page is challenging, especially for non-experts. It's also easy for novices to miss critical aspects of design ideation, such as stakeholders, consequences, and goals, when going through the design process ~\cite{macneil2021framing}. For these reasons, it is possible that additional scaffolding may be needed, especially for non-experts.

\subsection{Templates help novices at the cost of flexibility}

Although digital whiteboards provide designers with an unlimitled canvas that facilitates the free-form curation of their design work, templates also provide many benefits, especially for non-experts. In design settings,  templates have been studied extensively~\cite{pandey2018docent, macneil2021framing, hui2018introassist, bauer2013designlibs, deschamps2019archives, yu2015visual}. Templates can help to focus attention on the important design aspects~\cite{macneil2021framing, hui2018introassist} and guide users through a predefined process~\cite{pandey2018docent}. These conceptual and procedural scaffolds are especially helpful for non-experts who can not rely on their expertise to guide them. However, researchers have also consistently highlighted the negative aspect of templates which is that they are too rigid and lack flexibility~\cite{hui2018introassist, macneil2021framing}.


Currently, most commercial online whiteboards offer some form of templates that are either provided by the platform or sourced from their end-user communities. Despite the  prevalence of templates for digital whiteboards, few researchers have explicitly studied them. For example, Yarmand et al. use a template to support workshop participants, but don't explicitly evaluate the use of the template~\cite{yarmand2021facilitating}. Other researchers have studied digital whiteboards, but focus primarily on how creatives do their design work in the absence of templates~\cite{lupfer2016ideamache, linder2015beyond, kerne2017strategies}. This study sheds light on how templates are used in a free-form whiteboard environment to support collaborative design work.



    

\begin{figure*}
    \centering
    \includegraphics[width=\linewidth]{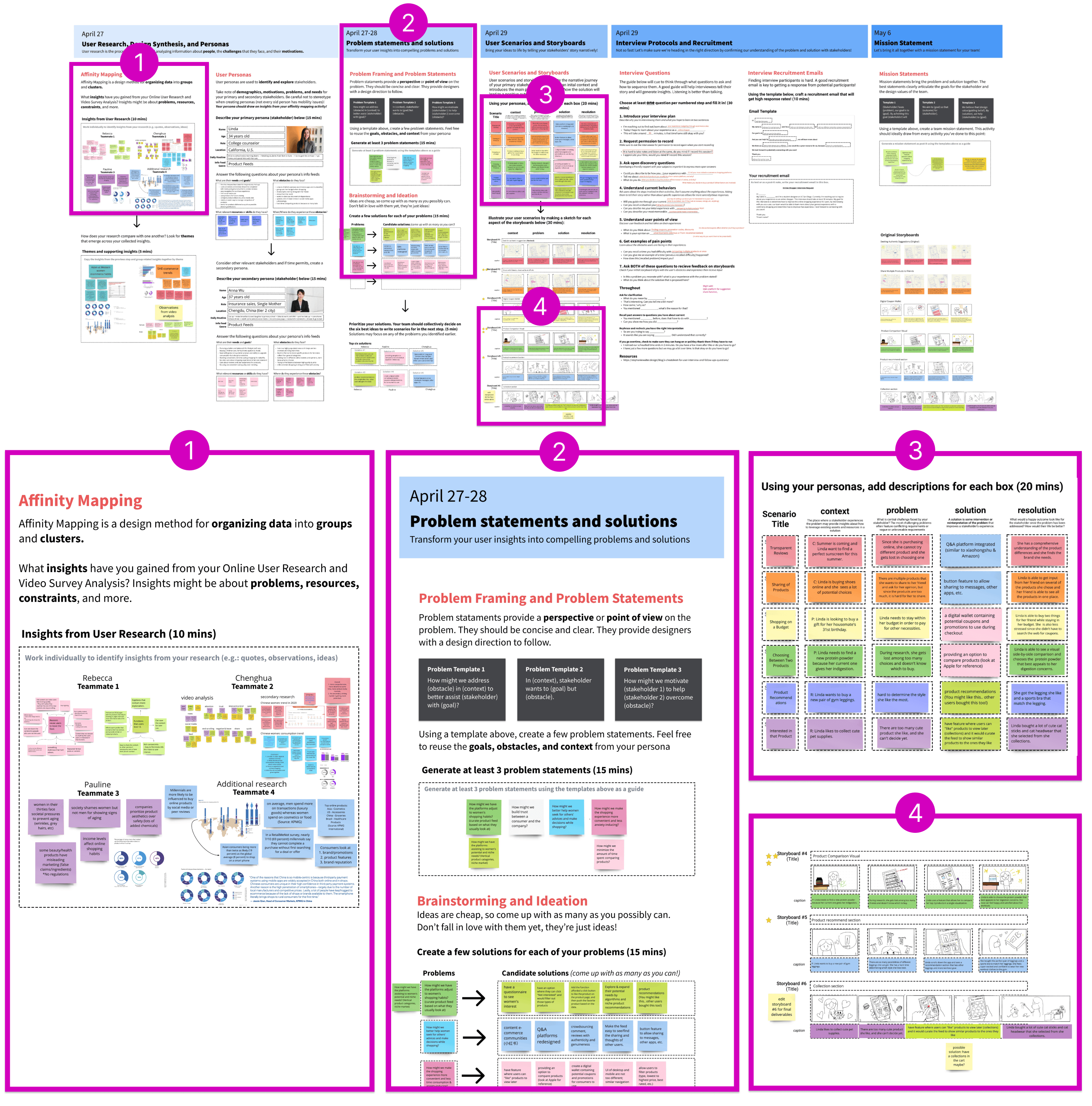}
    \caption{
    A template from the introductory design course filled in with students' colored sticky notes, images, and text. 1) Students were expected to add and organize sticky notes from user research, but in addition to this expected behavior, they added data, text, images, and created mind maps. 2) Panels focused on specific topics with instructions to guide contributions. Students additionally used colors to highlight themes across these activities. 3) Smaller input frames constrained what students contributed. 4) Students spontaneously used sticky notes to annotate and star emojis to vote for their favorite storyboards. 
    }
    \label{fig:design100-template}
\end{figure*}

\section{A Study of Template Usage in a Free-form Online Whiteboard} 

We hosted a multi-week design workshop as part of an introductory design class at the University of California, San Diego. The goal of this study was to better understand the potential for online collaborative whiteboards --- particularly the use of structured templates within the free-form environment --- to support design work. We focused on two main aspects: 1) a content analysis of the digital whiteboard, and 2) a post survey to understand participants' perspectives.

\subsection{Context and Participants} 

We recruited a faculty member at the University of California, San Diego (UCSD) to use online collaborative whiteboards for an introductory design class consisting of 114 students in Spring 2021. The course focused on developing interactive experiences through human-centered design processes such as need-finding, prototyping, and user testing. To support their design work, we developed a learning activity that guided them through their course design work. Each student filled out a consent form before participating in the activities and received one bonus point for completing the post survey. This research was approved by the Institutional Review Board (IRB) committee at UCSD.

\subsection{Workshop Overview} 


This workshop was designed to support students over the course a of a two-week design sprint. The workshop was hosted using Miro, an online whiteboard platform which provides a 2D canvas for students to add sticky notes and text. Miro provides affordances for adding, moving, and deleting these sticky notes. It also provides collaboration features like commenting and an awareness of collaborators' cursor locations on the canvas. Teams of 3-4 students used this Miro platform during and outside of class time to support their design activity. Over the course of four class sessions, teams entered into Zoom breakout rooms to complete different stages of activities in a provided activity template. An example of a template that has been filled in by a team is presented in Figure~\ref{fig:design100-template}.

\subsection{Workshop Template Design}



Most students in the class had limited prior design experience and the workshop was designed to provide these students with a structured process to follow that aligned with the course content. To ensure that the workshop fit the students' experience level and met the intended learning outcomes of the course, we co-designed the template with the instructor. The goal was to embed the instructor's expertise into the template to provide an expert design process that non-experts could follow with limited facilitation from the instructor. 

The activity template guided students to generate design artifacts (user research insights, personas, problem statements, storyboards, and interview protocols). Students were graded based on these design artifacts and a final presentation, but students were not required to use the online collaborative whiteboards and the provided template.  

The template consisted of a series of vertical panels laid out side-by-side horizontally with design activities stacked vertically within each panel (shown in Figure~\ref{fig:design100-template}). During each class period, students worked through a series of related design activities in a single panel by moving from the top of the panel to the bottom. Each panel covered a different part of the design thinking process. In addition to the blank template, a worked-out example was placed below the template which showed the template filled out with content from a different topic. For each activity within the panel, there was a title (e.g.: ``brainstorming and ideation''), corresponding guidance (e.g.: ``ideas are cheap, so come up with as many as possible. Don't follow in love with them just yet, they're just ideas!.''), and instructions about what contributions were expected (e.g.: ``create a few solutions for each of your problems (15 mins)''). This scaffolding was intended to guide students' design activity. 

\begin{figure*}
    \centering
    \includegraphics[width=\linewidth]{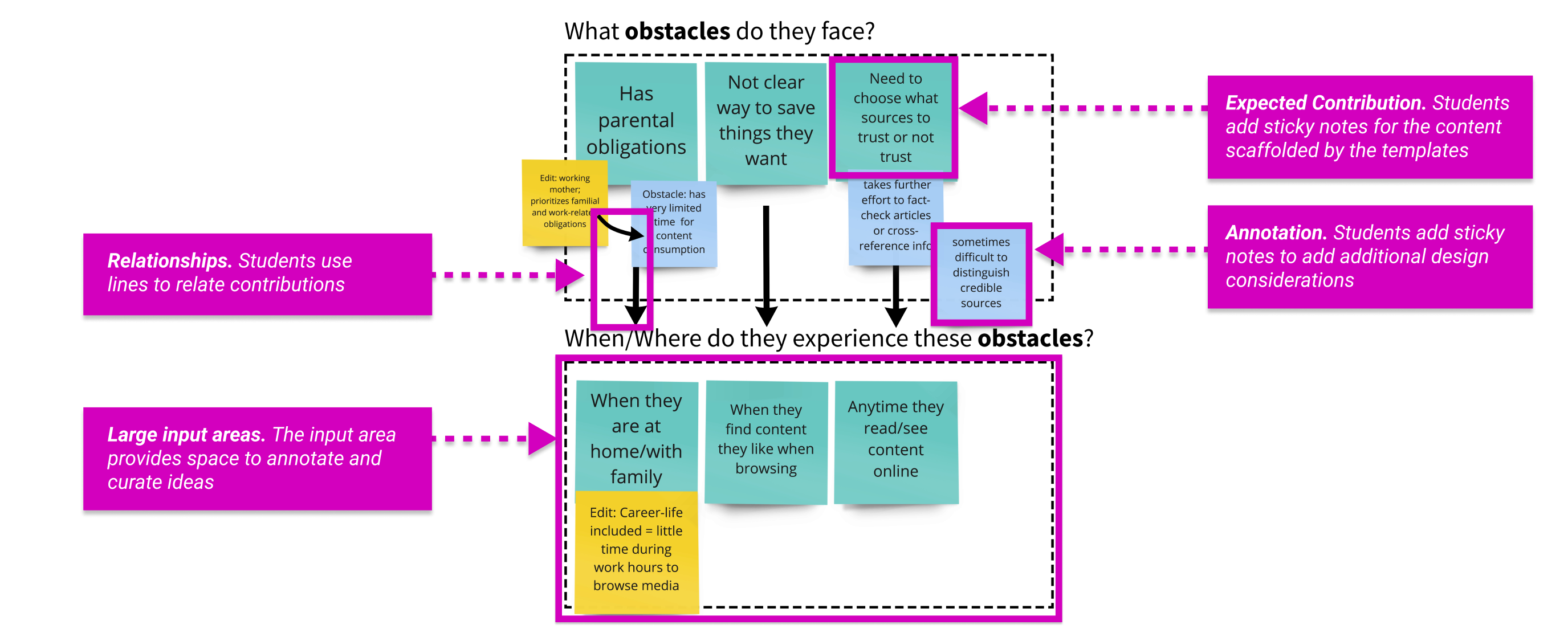}
    \caption{Free-form Templates. The templates included both smaller and larger input areas which were intended to guide user effort. In this example, the large input areas were both reappropriated to annotate and relate contributions. These free-form curation behaviors, which happened within and outside the input areas, are why we will refer to these templates in the discussion as `free-form templates.'} 
    \label{fig:free-form-templates}
\end{figure*}

The template included dotted boxes, which we refer to as \textit{input frames}, where students could add their design contributions. The design goal for providing frames with varying sizes was to implicitly indicate the amount of content students should contribute so that students would know how to allocate their effort. For example, brainstorming input areas were designed to be larger to accommodate more ideas. Although we anticipated that students would only make they types of contributions that the template scaffolded, the results of our study will show how these large areas were often reappropriated by students to support additional free-form curation that went beyond the contributions that the template scaffolded. Figure~\ref{fig:free-form-templates} provides an example of an input frame with free-form curation behaviors. Furthermore, we expected that students would work exclusively within these dotted areas despite having an unlimited space in the canvas outside the template.


\subsection{Methodology}

The goal of this study is to better understand how the free-form environment of a digital whiteboard along with templates affected participants' design processes and their resulting design outcomes (i.e.: design artifacts). To understand our participants' process and outcomes, we collected survey data about students' experiences along with images of the digital whiteboards that they used to complete their course design work. We analyzed this data with a mixed-methods approach that combined content analysis of the whiteboards with a thematic analysis of survey data.  
\fb{


\subsubsection{Whiteboard Content Analysis} 

To better understand students' process and outcomes, we analyzed students whiteboards using a mixed-methods content analysis methodology. Content analysis is a common method for analyzing the outcomes of a design process (i.e.: design artifacts). However, we used a modified content analysis method, inspired by design archaeology~\cite{chandra2019design} which enables us to infer things about the design process by analyzing the final design artifact. This design archaeology approach investigates aspects such as the appearance and function of an artifact, intended and unintended consequences of the artifact's usage, and analytical dimensions such as its symbolism and production context in order to understand a designer or user's intentions~\cite{chandra2019design}. As an example in our study, when students completed additional work outside of the template, we can infer that they temporarily stopped following the prescriptive design process provided by the template. Sometimes, we can further infer from the type of work they chose to complete why the template guidance was insufficient for that team.  Similar methodologies have been used in the past to analyze design artifacts and processes ~\cite{do2000intentions, identifying2019hu, reading2020changarana}.  For example, Do et al. previously used a similar method to understand the intention of designers by analyzing their sketches~\cite{do2000intentions}.  Hu et al. analyzed video and audio recordings from a design activity to identify divergent design thinking~\cite{identifying2019hu}.  Chang-Arana studied the ability to infer a user’s thoughts and feelings from a set of needfinding interviews~\cite{reading2020changarana}.

When studying the design process, researchers often use a participant observation methodology which can include verbal protocol analysis. These methods enable the research to understand the design process by observing designers and collecting verbal and visual data about their design process. We opted against this approach because our study was conducted in a highly naturalistic setting, where students are doing work both during class time and at home, making protocol analysis a less feasible method.

}

\subsubsection{Survey Data Analysis} 

Of the 114 students in the class, we collected 65 survey responses and Miro whiteboards from 37 teams. We applied inductive thematic analysis to the survey responses and whiteboards to investigate our key question around how and whether teams used the static templates within the freeform environment of an online collaborative whiteboard. The post design sprint survey included closed response questions about students' experiences with the whiteboards, templates, and teamwork. We also included open response questions about students’ experiences and any challenges they faced. We also conducted a content analysis of all 37 teams' whiteboards which we downloaded as vector images. Two researchers reviewed these whiteboards using an inductive coding method~\cite{saldana2015coding} to investigate overarching structures and behaviors that were common across the whiteboards. We also focused on identifying unique ways that the whiteboards were being used.

\section{Findings}

Out of a total of 37 teams, 34 used the whiteboards. The three teams that did not use the whiteboards had become inactive by this point in the semester. For the remaining 34 teams that used the whiteboards, our content analysis showed that all of them included both text and sticky notes posted by students. Three teams used the free-form handwriting feature to enter text in their whiteboards. In the following subsections, we present quotes from students' survey responses and our content analysis of the whiteboard templates. 

\subsection{Templates effectively scaffolded design work and aided collaboration}

\subsubsection{Templates gave teams a starting point}

Some participants highlighted the importance of having a starting point for their creative work: 

\begin{quote}
     ``Templates helped us get started on the project ... It really helped with the planning process.'' (P37)
\end{quote}

The worked-out example template was also helpful for participants by giving them an example of how to complete the activities. P17 said that their team, ``often referred to the examples to see if we were on the right track.'' The examples also served as a (re-)starting point when they got stuck:

\begin{quote} 
    ``We used the template as a reference often, and found it useful as examples to what our work should look like.'' (P8)
\end{quote}

The examples and instructions in the template as well as visual cues such as the size of the dotted input areas conveyed to students what was expected of their work, reducing confusion:

\begin{quote}
    ``[The templates] provide guidelines of what is it expected from the team.'' (P24)
\end{quote}

Students expressed how templates not only provide a starting point, but also helped teams create intermediate design artifacts that would be useful for later stages. Participants (5/65) expressed that the template helped them to connect one step to another.

\begin{quote}
    ``I think a benefit was being able to see what steps we were on and how each step follows the next. This makes everything cohesive and easy to follow.'' (P33)
\end{quote}

As a starting point, multiple students (12/65)  explicitly described how the template made the design process more beginner-friendly. However, a few students (5/65) also described usability issues, features that were not discoverable, or a learning curve. For example, 

\begin{quote}
    ``In the beginning the functions were hard to utilize because I never utilized this platform before, however after playing around I got more comfortable.'' (P33)
\end{quote}







\subsubsection{Templates helped teams coordinate around a shared process}


When asked to describe their experience, many students (28/65) explicitly described that the templates gave them a necessary process to follow throughout the workshop. For example,



\begin{quote}
    ``The template made it easily to stay on track with the stages of the project. Everything was organized into sections that built off each other.'' (P46)
\end{quote}

The template provided a shared process that enabled teams to develop common ground~\cite{clark1991grounding, fischer2005knowledge} related to their work. This was especially helpful when working with new people or with people that have different design skills and experience:

\begin{quote}
    ``[The whiteboard and template were] especially useful when working with people I do not know, because it made a base template that we could all follow regardless of experience level.'' (P13)
\end{quote}

While the process provided by the templates was helpful to keep participants on track and make effective progress through the workshop, the templates also served as a \fb{real-time dashboard} which helped the team assign work to team members, monitor each team member's progress, and empower team members to give timely feedback. For instance, P21 described how their team used the whiteboard to monitor individual team members' progress:

\begin{quote}
``... [we got] real time updates of what each team member was doing as we were all working together.'' (P21)
\end{quote}

This real-time awareness also enabled the team members to coordinate work. As P31 said, ``it was very easy to organize our work and assign work to each team member.'' P24 described how monitoring his team prompted him to provide feedback, ``[it was] nice to see what other people wrote so I can give feedback'' (P24). P17 wrote that the templates helped with planning:
\begin{quote}
``It seemed easy for all team members to follow along and understand what we're doing and what we plan to do the next day.'' (P17)
\end{quote}



Finally, the template appeared to provide students with a visual summary of their work and progress:

\begin{quote}
``It was really helpful to have a cohesive and visual representation of our work and steps along the way. It was easy to see what we had to complete'' (P30)
\end{quote}


\subsubsection{Participants often referred to information in previous steps}

Participants described how the whiteboard was a constant reference point and resource throughout their design process:

\begin{quote}
    ``The whiteboard was our home base for working on the project up until we started drafting our presentation, because it was easy for us to expand on our work since our previous work was all in one place. The templates kept us on track and helped us understand how different stages in the project interacted and influenced each other.'' (P48)
\end{quote}

Participants often looked back at their previous steps and design contributions to help get started on subsequent steps:

\begin{quote}
    ``I also like that we used Miro throughout the entire project because it was very easy to track progress and look back to previous steps when coming up with ideas further down the line.'' (P53)
\end{quote}

Three participants used `flow' to describe how the templates guided them through design activities. For instance, P34 said:

\begin{quote}
    ``...we were able to go back and see the previous steps such as problems, solutions, storyboards etc. The whiteboard helped us organize information in a clear way and we could just follow the flow.'' (P34)
\end{quote}

One of the main benefits participants (10/65) shared for reviewing the templates was the ability to organize and connect ideas: 

\begin{quote}
    ``It was nice to synchronously view each other’s inputs...it helped us visualize our progress and easily connect one another’s ideas/thoughts.'' (P28)
\end{quote}

\subsubsection{Despite the benefits, a minority of students believed templates were restrictive}

The templates were a helpful scaffold to get teams started, guide them through a process, and support reflection and reinterpretation of their prior activity. However, some participants also cited shortcomings of structuring a design workshop with templates. For instance, several students (16/65) mentioned that the templates were rigid and that teams may have their own preferences for how they wish to carry out the design process that deviates from provided template. P24 said that the workshop template ``forces the user to follow the structure.'' P48 elaborated on this idea by implying possible pros and cons associated with template use: 

\begin{quote}
``Although the template was useful for this particular project, I could see how having so many constraints could limit how a person uses the whiteboard.''
\end{quote}


\begin{figure*}[!ht]
    \centering
    \includegraphics[width=\linewidth]{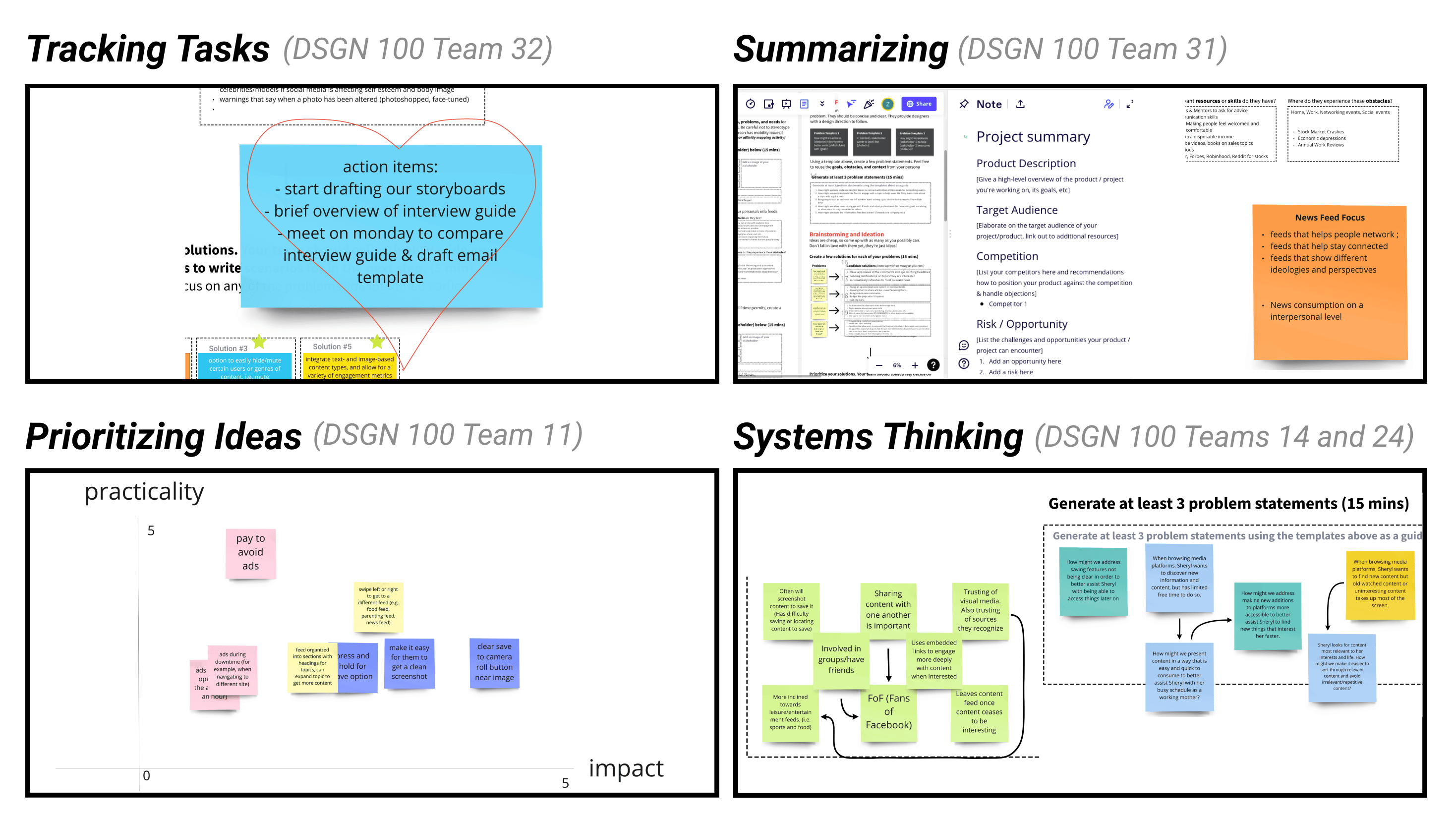}
    \caption{The content analysis shows how participants worked within and outside templates to provide additional structure and flexibility. Participants used sticky notes to track tasks, both sticky notes and the sidebar to summarize ideas, and open spaces to do additional activities like priority matrices. They also used arrows, spatial grouping, and sticky note color to convey additional meaning. In each of these cases, teams went beyond what was scaffolded by the templates. }
    \label{fig:flexibility}
\end{figure*}

\subsection{Participants created flexibility inside and outside the templates}


We observed in our content analysis many ways that participants created additional structure and flexibility both within and outside the templates. Figure~\ref{fig:flexibility} shows a few examples of the ways that participants worked outside and inside the lines. The following subsections present the primary findings from our content analysis.


\subsubsection{Participants further curated their contributions within the input frames}

Within the templates, where students were instructed to add sticky notes, we observed how teams added additional meaning by clustering related sticky notes together (15/34), color coding them (31/34), or creating arrows to represent relationships between sticky notes. Team members also used color to differentiate their own contributions from others. This behavior may demonstrate a need for attribution or may be a necessary method for coordinating design activity and synchronizing their efforts. A few teams (2/34) also used the comment function to coordinate or to provide feedback, for example, one student left the following comment to their team members: 

\begin{quote}
``It would really help set up this scenario if you made it more specific by putting in a specific topic \textit{redacted name} is looking for.'' (Team 3)
\end{quote} 

In addition, three teams used sticky notes as comments to communicate at a meta-level about their design work. For example, on Team 6 there was a sticky note suggestion for the rest of the team, ``Might add: Q\&A platform for suggestion share function.'' 

\subsubsection{Participants used the area outside the template for ad hoc design work}

The large 2D canvas provided by Miro offered students ample space outside of the template for additional design activity. We expected students would work explicitly within the template; however, based on our content analysis, we  observed that a little under half of the teams (16/34) completed design work outside the designated boundaries of the template. For example, one team found open space in the whiteboard to create a moodboard. Another team created priority matrices (see Figure~\ref{fig:flexibility}, Team 11). In addition to completing these self-scaffolded design activities, we also observed students using the space outside the template for meta-design work. For example, students used the Miro Sidebar, comments, and sticky notes to summarize their design work and assign tasks.  

\subsubsection{Participants created hyperlinks within the template to external work}

While the template was designed to provide students with the scaffolding they needed to complete their design work, we observed a few teams (4/34) that completed at least one of the activities in an external tool and linked to that tool in the whiteboard. For instance, some students chose to do storyboarding in a specialized tool with storyboard features, while others opted to write their interview questions in Google Docs and link to it in the provided template. These behaviors suggest that the Miro board and template served as a hub for collaboration.

\section{Discussion}


In this study, we investigated how students interacted with structured design templates within Miro's free-form online collaborative whiteboard. Our results provide novel insights about how students used the templates in this free-form environment. 


\subsection{Digital whiteboard templates provide important benefits for non-experts}

Based on our content analysis of the whiteboards and student survey data, we observed evidence that templates provide an essential starting point and re-starting point for collaboration. The template appears to provide initial creative constraints~\cite{moreau2005designing, sternberg1999concept} that help students avoid fixation related to the limitless possibilities of the whiteboard and `blank page' effect~\cite{joyce2009blank}. In addition to this starting point, templates also model an expert process that is easy to follow and which `grounds'~\cite{clark1991grounding, fischer2005knowledge} the team's communication, even between designers who have not worked together previously or who have different levels of design experience. Templates provided a `visual representation of progress' and supported reflection on previous work that had been completed. 

Students talked about frequently revisiting and reflecting on previous work to inform future steps and design activity. Providing students with explicit opportunities to revisit their work in new contexts could promote creativity recombination~\cite{wilkenfeld2001similarity} and reflection. Similarly, students talked about how the template provided them with a `visual representation of progress' and a `visual representation' of their completed work. Finally, students talked about using the worked examples below the template to help guide their next steps. 

Aligned with prior research~\cite{macneil2021framing, hui2018introassist}, we found the templates to be restrictive, with 24.6\% of students mentioning that they would have preferred a different template or that the template restricted them in some way~\cite{macneil2021framing, hui2018introassist}; however, we also learned that participants were largely positive about their experiences with the templates. We observed how participants created additional flexibility and structure by working inside and outside the templates. 

Participants used sticky notes, comments, and completed additional design work outside the template to achieve additional flexibility. Within the template input areas, participants created additional structures with arrows to relate sticky notes to each other. They also used color and spatial grouping to add additional meaning and structure to their free-form input areas. Our findings reveal how free-form online whiteboards might make the experience of using templates feel less rigid. This suggests that templates and free-form curation~\cite{kerne2017strategies, linder2015beyond, lupfer2016patterns} can be synergistic.

\subsection{Free-form Templates: free-form curation in a structured design process}



Previous research has demonstrated the potential for free-form curation to help designers discover emerging relationships and design concepts ~\cite{kerne2017strategies}. Researchers also discovered patterns of visual thinking that emerged organically when students organize their ideas through free-form curation~\cite{lupfer2016patterns}. For example, students spatially arranged and grouped elements to creatively explore and creatively associate concepts to form new understandings. In those studies, students were able to explore divergent paths and they layered content to create visual hierarchy and think at multiple levels of abstraction. Yet, free-form curation relies on chance and on the user's ability to curate their work. While templates can provide guidance, they have been criticized for being too restrictive~\cite{macneil2021framing, hui2018introassist}. The results from our study, suggest that using templates in a free-form environment has the potential to be a synergistic approach, creating a bridge between these two previously disparate concepts. In addition to the traditional scaffolding benefits provided by the templates in our study, we also observed that students created additional structure and flexibility for themselves while working both inside and outside the templates. 

We coin the term \textit{free-form templates} to refer to templates that are used within a free-form environment and feature large input areas to guide user contributions. These templates allow for free-form input while also providing the traditional structure and scaffolding associated with templates.

\subsubsection{Free-form curation inside and outside the templates} 

Templates have been shown to help non-expert designers~\cite{bauer2013designlibs, hui2018introassist, macneil2021framing}. Yet, templates have also been criticized for being overly restrictive, which may be problematic for creativity. \textit{Free-form templates} allow free-form input with a large and zoomable space for contribution within the template. The large open area provided students with an opportunity to creatively organize and group design concepts to discover new relationships, meaning, and ideas. Students also have unlimited space beyond the template for creative exploration. 

In our study, when students used the \textit{free-form templates}, we observed patterns of free-form curation~\cite{kerne2017strategies} where students worked both inside and outside the provided template to create additional structure and flexibility. Outside the template, students completed supplemental design work such as creating moodboards and priority matrices. They also added comments and sticky notes as meta-design-work to coordinate their team efforts and to reflect and summarize their design work. Inside the template, we observed how students color coded sticky notes, spatially organized and grouped them, and created arrows to explore relationships between design concepts. 

Finally, we observed evidence of students linking to other tools within their digital whiteboard environment. Typically, this took the form of replacing scaffolded activities with external tools that specialize in supporting that design activity. These observations align with recent calls~\cite{fox2020towards, borowski2022varv} to break through the walled gardens and information silos that require users to switch between tools in a toolbelt~\cite{sumner1997evolution} or stitch the seams between the platforms that they use~\cite{dailey2017social}. These observations also add to an ongoing conversation about how designers sometimes need to play the role of seamster to bring together bits and pieces of ideas across a larger design platform~\cite{macneil2021finding, macneil2018designing}. 

\subsubsection{Free-form templates balance flexibility with guidance}

Based on our study, \textit{free-form templates} appear to provide flexibility while effectively scaffolding students through their design process. Borrowing terms from free-form curation, the template itself provided students with a `path' to follow, but we observed cases where students diverged from the given path into unexpected spaces. Templates also help organize processes and details by providing a `visual hierarchy' of high-level design activities with low-level details. The path and visual hierarchy helped teams to frequently refer to information in previous steps and associate concepts across activities and input areas, suggesting that different input areas are not segregated by the template but rather connected as a `visual semantic whole' to support further ideation inside and outside the template. These paths, hierarchies, and perceiving the work as a visual semantic whole are hallmarks of visual patterns of free-form curation~\cite{kerne2017strategies}.

\subsection{Future Work: Free-form templates capture \textit{Design Data}}



Previous researchers have identified many challenges associated with documenting
the design process which include additional effort for the designer, time taken away from design work, and potential for interrupting the designer's creative flow~\cite{tseng2014documentation, bardzell2016documenting, milara2019document}. To reduce the extra work required by documentation, researches have explored techniques to ``document while doing''~\cite{milara2019document}. However, in the case of free-form templates ``doing \textit{is} documenting.'' The template represents an evolving design process and by capturing key decisions. For example, in our template, there were spaces for teams to add stakeholders, problems, and solutions. By augmenting these whiteboards with event-based data tracking, it might be possible to collected these design decisions for each team and even generate reports that participants could use for onboarding new team members, reflecting on previous work, and communicating with stakeholders. 


\subsubsection{Frames effectively collect and label data}

In our study, we observed some initial evidence that free-form templates can effectively capture participants' design contributions as structured \textbf{design data}. Our templates contained large dotted boxes which we call \textit{input frames} where students could add their design contributions. Participants in this study correctly placed their sticky notes within these input frames with one exception, where one participant put multiple sticky notes around an input frame and only the sticky note inside the frame was captured. These input frames made it possible to easily capture and store design data from the whiteboard into a database with a label that corresponded to the input frame in the template from which the data was collected. This allowed us to know whether the sticky note referred to a stakeholder, obstacle, solution, or some other design data type. \fb{ 

\subsubsection{The potential uses of design data}

Previous creativity support tools often leverage labeled data to make personalized recommendations~\cite{maher2017encouraging, sielis2011context} or to support faceted search in the context of design galleries~\cite{kang2018paragon, siangliulue2016ideahound, macneil2021probmap, lee2010designing, masson2014web, dziubak2018prism, herring2009getting, mendels2011freed}. In prior work, this typically this requires substantial human labor through crowd labeling~\cite{hope2017accelerating, kang2018paragon}, algorithmic approaches that can extract limited features~\cite{macneil2021probmap} or by identifying visual groupings to infer meaning~\cite{jain2021recognizing}. Our research shows how templates provide the context necessary to accurately capture heterogeneous design data in real-time by leveraging input frames within the template. 
Further research could illuminate how this design data should best be leveraged by creativity support tools. For example, presenting teams with automated summaries of their work and the work of others may provide value, but there is risk that design data can be decontexualized without explicit pointers or \fb{information scents~\cite{pirolli2000effect}} that preserve the context and provenance of the design data. 

Finally, in addition to obtaining labeled design contributions, input frames may also provide additional context about ontological relationships between the design data. For example, frames might collected specific data types such as \textit{stakeholders, problems, goals, solutions}, and other design information. These data types can theoretically be related to each other through ontological relationships. For example, $(Stakeholder) - [ :hasA ] -> (Problem)$. This begins to show the potential for design data to be used to construct a design data model. Future work could explore opportunities to create multiple representations of the underlying design data. For example, the WritLarge system shows how users can flexibly transition between equivalent representations---such as between handwriting and text~\cite{haijun2017writlarge}. In our case, this might be transitioning between a stakeholder, goal, and obstacle and a problem statement that contains that information. 

}





\section{Conclusion} 
\fb{
In this paper, we conduct a mixed-methods study of how design students use templates in a free-form digital whiteboard environment. We show how templates can help guide students through a design process; providing them a starting point, procedural guidance, and a visual representation of their work that helped coordinate each team's design activity and support reflection on their design work. We also introduce the concept of \textit{free-form templates} which provide students with structure and scaffolding while also providing additional flexibility that is necessary for creative work, but that is not common in traditional templates. We discuss the implications that these free-form templates have for design documentation and the collection of design data, while still providing the traditional benefits associated with templates. 
}


\begin{acks}

We would like to thank the students and instructional team for the introductory design class at UCSD. We would also like to thank the reviewers for their feedback which helped to improve the final version of the paper. 

\end{acks}

\balance
\bibliographystyle{ACM-Reference-Format}
\bibliography{sample-base}

\end{document}